\documentstyle[aps,preprint,prl]{revtex}

\begin{document}
\bibliographystyle{plain}
\title{Effect of dissipation on the decay-rate phase transition}
\author{Soo-Young Lee$^a$, Hungsoo Kim$^b$, D.K.Park$^a$, Chang Soo Park$^c$,
Jae Kwan Kim$^b$}
\address{$^a$ Department of Physics, Kyungnam University,
Masan, 631-701, Korea.\\
$^b$ Department of Physics, Korea Advanced Institute of Science and
Technology, \\Taejon, 305-701, Korea.\\
$^c$ Department of Physics, Dankook University, Cheonan, 330-714, Korea.}
\date{\today}
 \maketitle

\tightenlines
 \begin{abstract}
A general condition for sharp transition of decay rate from quantum
to thermal regimes is derived in dissipative tunneling models
when position-dependent mass is involved.
 It is shown that the effect of dissipation in general changes the order of
 the phase transition.
 Especially, for the models with constant mass the Ohmic dissipation 
enlarges 
 the range of parameters for first order phase transitions.
In the case of second-order phase transition the Ohmic dissipation
suppresses the decay rate near the transition temperature($T_c$). 
For the super-Ohmic case the dissipation yields an opposite effects to 
the Ohmic dissipation within exponential approximation.
\end{abstract}
\newpage
\section{Introduction}
It is well known that the decay of the metastable states at zero temperature
is determined by pure quantum tunneling process whose dynamics is
described by classical configurations called bounce in Euclidean
 space\cite{co77}.
Since, however, the decay rate is determined at high temperature by thermal
activation which  corresponds to classical configuration called
sphaleron\cite{kl84}, there exists some temperature $T_c$ at which the
transition from classical- to quantum-dominated decay occurs. This phase
transition problem was firstly discussed by Affleck\cite{af81} within
quantum mechanics. He demonstrated that under certain assumptions
for the shape of the barrier the transition between the thermal and quantum
regimes is dominated by solutions with a finite period in the Euclidean time
that smoothly interpolate between the zero-temperature bounce and the
static high-temperature sphaleron. The transition is thus a second-order one.
Chudnovsky\cite{ch92}, however, has shown that the type of the phase
transition in the crossover from thermal activation to thermally assisted
quantum tunneling is completely dependent on the shape of the potential
barrier. He also has shown that the order of the phase transition is easily
conjectured by $P$-vs-$E$ graph, where $P$ and $E$ are Euclidean period
and energy, repectively. The sharp crossover between the thermal and
thermally assisted tunneling occurs when $P=P(E)$-curve possesses
a minimum at $E=E_c$, which is different from the energy of sphaleron
solution.
Based on Chudnovsky's observation the sharp first-order transitions are
found at spin tunneling systems with\cite{le98,ch97} and without\cite{li98}
 external magnetic field.

Recently, a sufficient criterion for the first-order phase transition in decay
problem of the metastable state is obtained by carrying out the nonlinear
perturbation near the sphaleron solution in the two-dimensional string
model\cite{go97}. Inspired by spin-tunneling problem, the result of Ref.\cite{go97} is
subsquently
extended to the quantum mechanical model when mass is position-dependent\cite{mu98}.

The purpose of this paper is to derive a general criterion for the sharp
first-order phase transition in the quantum mechanical tunneling models
when the position-dependent mass and dissipation are involved.
We first consider a system with  Ohmic dissipation in which the spectral 
density $J(\omega)$ is linearly proportional to frequency, i.e.,
 $J(\omega)=\alpha \omega$ where $\alpha$ is the dissipation coefficient.
 In this case Caldeira and Leggett already derived the effective
action in their seminal paper\cite{ca83}. We will use their result for
the derivation of the criterion. 
The extension to the case of super-Ohmic dissipation where 
$J(\omega)=\alpha \omega^3$ is also examined. In the super-Ohmic case 
a counter term corresponding to the deformation of potential due to
environment is introduced.
The absence of the counter term in the Ohmic case is due to 
our use of the result of Ref.\cite{ca83}, in which a proper counter
term is already used for the Ohmic dissipation.
Since we work at Euclidean space, we use a prescription for the dissipation
 term which does not break 
the time-reversal symmetry in the both Ohmic and super-Ohmic cases.
Carrying out the nonlinear perturbation we will derive the general
condition for the first-order phase transition in the dissipative quantum
mechanical system when mass is position-dependent in Sec.II.
 It is found that the effect of dissipation is only deformation
of the eigenvalues for the temporal fluctuation operator $\hat{l}$ defined 
in this section.
To get some physical intuition we will apply this general criterion
to the simple quantum mechanical models in Sec.III.
 Brief conclusion and the direction of the future research in this field
will be given at the final section.

\section{Criterion of first-order phase transition for dissipative
quantum systems}
The effect of dissipation on quantum tunneling is investigated
in Ref.\cite{ca83} by introducing the infinite number of oscillators as an
environment and assuming the linear coupling between the environment and
system. We follow their formalism to obtain the effective action for
the dissipative system.

The effective action in Euclidean space is given as
\begin{equation}
S[q(\tau)]=\int_{0}^{T} [\frac{1}{2} M(q) \dot{q}^2 + V(q) +\delta V(q)] d\tau
+\frac{1}{2} \int_{-\infty}^{\infty} d\tau'\int_{0}^{T} d\tau
\gamma(\tau -\tau')\{ q(\tau) - q(\tau')\}^2,
\label{effaction0}
\end{equation}
where 
\begin{equation}
\gamma(\tau - \tau') \equiv \frac{1}{2\pi} \int_0^\infty J(\omega)
 e^{-\omega \mid \tau - \tau' \mid} d\omega.
\end{equation}
Here, 
the mass of quantum system
is generally taken as position-dependent which is motivated from
spin tunneling models\cite{le98,ch97,li98}. The final non-local term in
Eq.(\ref{effaction0}) represents the effect of
 dissipation.
It is emphasizing to note that the counter term $\delta V(q)$ is introduced
in Eq.(\ref{effaction0})\cite{fu92}.
In fact, the above effective action without $\delta V(q)$ has been derived
 in consideration
of a counter term which cancels divergence in the case of Ohmic dissipation.
However, for the cases of non-Ohmic dissipation new divergence is expected to appear,
so that appropriate conter term may have to be needed.
Shortly, it will be shown that  this is a generic case.

Now, we take the spectral density generally as
\begin{equation}
J(\omega) = \alpha \omega^n
\end{equation}
where $n$ is an positive integer.
Then, the effective action can be written as
\begin{equation}
S[q(\tau)]=\int_{0}^{T} [\frac{1}{2} M(q) \dot{q}^2 + V(q)+ \delta V(q)] d\tau
+\frac{\alpha \Gamma (n+1)}{4\pi} \int_{-\infty}^{\infty} d\tau'\int_{0}^{T} d\tau
\frac{\{ q(\tau) - q(\tau')\}^2}{\mid \tau - \tau' \mid^{n+1}},
\label{effaction2}
\end{equation}
where $\Gamma(n)$ is the Gamma function.
From this effective action, the equation of motion is given by
\begin{equation}
M(q)\ddot{q} + \frac{1}{2} M'(q) \dot{q}^2
-\frac{\alpha \Gamma (n+1)}{\pi} \int_{-\infty}^{\infty} d\tau'
\frac{q(\tau) - q(\tau')}{\mid \tau - \tau'\mid^{n+1}} = V'(q) +\delta V'(q),
\label{eqmotion}
\end{equation}
where the prime denotes the derivative with respect to $q$. 
In the following subsections we will derive the criterion of the first-order
phase transition in the cases of Ohmic dissipation and super-Ohmic 
dissipation($J(\omega )=\alpha \omega^3$).

\subsection{The Ohmic dissipation}
In the case of Ohmic dissipation, the spectral density is linearly
proportional to frequency. The equation of motion is, thus, 
Eq.(\ref{eqmotion}) with $n=1$.
The dissipation term  can be rewritten in terms of the Fourier transform
partner of $q(\tau)$, i.e., $\tilde{q}(\omega)$, which makes the equation
of motion to be
\begin{equation}
M(q)\ddot{q} + \frac{1}{2}\frac{\partial M(q)}{\partial q} \dot{q}^2
-\alpha \int_{-\infty}^{\infty} d\omega
\tilde{q} (\omega) \mid \omega \mid  e^{i\omega \tau} = V'(q).
\label{eqmo}
\end{equation}
In the derivation of this equation we take a prescription for the dissipation
term
\begin{equation}
-\frac{\alpha}{\pi} \int_{-\infty}^{\infty} d\tau'
\frac{q(\tau) - q(\tau')}{( \tau - \tau' )^2 }
\,\,\,\rightarrow  \,\,\,
-\frac{\alpha}{\pi} \int_{-\infty}^{\infty} d\tau'
\frac{q(\tau) - q(\tau')}{( \tau - \tau' +i\epsilon) ( \tau - \tau' 
-i\epsilon)},
\end{equation}
which preserves the time-reversal symmetry,
and $\delta V'=0$ as mentioned before. 
Since, in fact, the dissipation breaks the time-reversal symmetry
in real space-time, another prescription was used at Ref.\cite{fu92}.
However, in Eclidean space the dissipation does not yield damping motion,
which justifies our choice of the prescription.

Now, following Ref.\cite{mu98}, let us determine the type of transition by
 expanding the equation of motion at the sphaleron solution $q_s$ as
\begin{equation}
q(\tau) = q_s + a\eta_1 (\tau)
\end{equation}
or equivalently
\begin{equation}
\tilde{q}(\omega) = \tilde{q}_s + a\tilde{\eta}_1(\omega),
\end{equation}
where $a$ represents an oscillation amplitude near sphaleron solution.
Since it is sufficient in determining the order of phase transition
to consider only the solutions near sphaleron, we can assume $a$ is
very small constant.
Substuting these expressions into Eq.(\ref{eqmo}) the equation of motion
within the first order of $a$ becomes
\begin{equation}
(\hat{l} -\hat{h} ) \eta_1(\tau) = 0,
\label{eta1}
\end{equation}
where operators  $\hat{l}$ and $\hat{h}$ are defined as
\begin{eqnarray}
\hat{l} &=& M(q_s)\frac{d^2}{d\tau^2} - \frac{\alpha}{2\pi} \int d\omega
\mid \omega \mid  e^{i\omega \tau}  \int d\tau 
e^{-i\omega \tau},       \nonumber      \\
\hat{h} &=& V''(q_s),
\end{eqnarray}
respectively.
In order to solve this equation, we take a trial solution\cite{go97,mu98}
\begin{equation}
 \eta_1(\tau) =  \cos \Omega \tau
\end{equation}
and equivalently
\begin{equation}
\tilde{\eta}_1 (\omega) = \frac{1}{2} (\delta (\Omega -\omega ) +
\delta (\Omega +\omega )).
\end{equation}
Using this trial solution the frequency $\Omega$ near the sphaleron
solution is obtained within
the first order of the amplitude $a$ as
\begin{equation}
\Omega_O^{(1)} = \pm \frac{1}{2} \left[ -\frac{\alpha}{M(q_s)} +
 \sqrt{(\frac{\alpha}{M(q_s)})^2 + 4 \omega_s^2} \right]
\label{Omega1}
\end{equation}
where
\begin{equation}
 \omega_s = -\frac{V''(q_s)}{M(q_s)}.
\end{equation}
Comparing $\mid \Omega_O^{(1)}\mid$ with $\omega_s$ which is
$\alpha \rightarrow 0$ limit of  $\mid \Omega_O^{(1)} \mid$,
it is easy to understand that the Ohmic dissipation reduces the frequency.
Since in the case of second-order phase transition the transition temperature
$T_c$ is determined by this  $\mid \Omega_O^{(1)}\mid$, i.e., 
$T_c = \mid \Omega_O^{(1)}\mid/2\pi$, this result implies that the Omic
dissipation decreases the transition temperature.
As shown in Fig.1, the decrease of $T_c$ means increase of action value
and, then, suppression of decay rate near $T_c$. We will, however, show that 
the super-Ohmic dissipation gives the opposite effect on decay rate
within exponential approximation.

Next order calculation can be conducted by taking
\begin{equation}
 q(\tau)=q_s+ a\eta_1(\tau) + a^2 \eta_2(\tau).
\end{equation}
Then, we find  equation of motion for $\eta_2(\tau)$ as
\begin{equation}
a ( \hat{l} -\hat{h} )\eta_2(\tau) = -( \hat{l} -\hat{h} )\eta_1(\tau)
+ a W_1(\tau),
\label{eta2}
\end{equation}
where
\begin{equation}
W_1(\tau) =
(\Omega^2 M'(q_s)+\frac{1}{2}V'''(q_s)) \cos^2 \Omega \tau
- \frac{1}{2}\Omega^2 M'(q_s) \sin^2 \Omega \tau .
\end{equation}
Since $(\hat{l}-\hat{h})$ is an hermitian operator, 
taking a scalar product with $\mid \eta_1>$ on both sides of
the equation of motion yields
\begin{equation}
a(l(\Omega)-h(\Omega))<\eta_1 \mid \eta_2 > =
-<\eta_1 \mid \hat{l}-\hat{h} \mid \eta_1 > + a <\eta_1 \mid W_1 >,
\label{eta2a}
\end{equation}
where
\begin{eqnarray}
l(\Omega)&=&-\Omega^2 M(q_s)-\alpha \mid \Omega \mid ,\nonumber \\
h(\Omega)&=& V''(q_s).
\end{eqnarray}
As usual perturbation theory, $<\eta_1 \mid \eta_2 >$ is zero.
Furthermore, the second term of the right-hand side in Eq.(\ref{eta2a}) is
 zero because of $\tau$-integration.
Then, the above equation(Eq.(\ref{eta2a})) is identical with Eq.(\ref{eta1}).
Therefore, within the second order of $a$
we can not find the variation of the frequency from $\Omega_O^{(1)}$,
i.e.,
\begin{equation}
\Omega_O^{(2)}=\Omega_O^{(1)}.
\end{equation}
Before calculating next order frequency $\Omega_O^{(3)}$, we would like
to evaluate $\eta_2(\tau)$ explicitly for later use.
This is achieved directly from  Eq.(\ref{eta2}), which yields
\begin{eqnarray}
\eta_2(\tau) &=& (\hat{l}-\hat{h})^{-1} W_1(\tau) \nonumber  \\
             &=& g_1 + g_2 \cos 2\Omega \tau ,
\label{etatwo}
\end{eqnarray}
where
\begin{eqnarray}
g_1 &=& - \frac{\Omega^2 M'(q_s) + V'''(q_s)}{4 V''(q_s)} ,\nonumber \\
g_2 &=& - \frac{3 \Omega^2 M'(q_s) + V'''(q_s)}
     {4[ 4 M(q_s)\Omega^2 + \alpha \mid 2 \Omega \mid + V''(q_s)]}.
\label{g1g2}
\end{eqnarray}

Now, taking a third order correction into $q(\tau)$ as
\begin{equation}
 q(\tau)=q_s+ a\eta_1(\tau) + a^2 \eta_2(\tau) + a^3 \eta_3(\tau ),
\end{equation}
and inserting the above expression into Eq.(\ref{eqmo}),
the equation of motion for $\eta_3(\tau)$ is straightforwardly obtained as
\begin{equation}
a^2 (\hat{l} -\hat{h} ) \eta_3 =-(\hat{l} -\hat{h} ) \eta_1
-a(\hat{l} -\hat{h} ) \eta_2 + aW_1(\tau)+ a^2 W_2(\tau),
\label{eta3}
\end{equation}
where
\begin{eqnarray}
W_2(\tau )& =& (\Omega^2 g_1 M'(q_s) + g_1 V'''(q_s))\cos \Omega \tau
 \nonumber \\
& &+(5 \Omega^2 g_2 M'(q_s) +g_2 V'''(q_s))\cos 2 \Omega \tau \cos \Omega
 \tau \nonumber \\
& &+\frac{1}{2}(\Omega^2 M''(q_s)+\frac{1}{3} V'''')\cos^3 \Omega \tau \\
& &- 2 \Omega^2 g_2 M'(q_s) \sin \Omega \tau \sin 2\Omega \tau \nonumber \\
& &-\frac{1}{2}\Omega^2 M''(q_s) \cos \Omega \tau \sin^2 \Omega \tau .
\nonumber
\end{eqnarray}
Scalar product with $\eta_1$ in Eq.(\ref{eta3}) yields an equation
\begin{equation}
-(l(\Omega) - h(\Omega))<\eta_1 \mid \eta_1>
+a^2 <\eta_1 \mid W_2 > =0
\label{lh}
\end{equation}
in which we can determine $\Omega_O^{(3)}$.
Since the first-order phase transition between thermal and thermally assisted
quantum tunneling regimes occurs when the period of solution near the
sphaleron increases with approaching to the sphaleron solution,
 we get a condition for sharp transition by  $\mid \Omega_O^{(3)} \mid >
\mid \Omega_O^{(1)} \mid $.
 This is identical to the condition that the value of
the left-hand side of Eq.(\ref{lh}) at $\Omega =\mid \Omega_O^{(1)}\mid $
 is negative.
Therefore, we finally obtain the criterion of first-order phase transition
as
\begin{equation}
\Omega_O^{(1) 2} [(g_1+\frac{3}{2}g_2) M'(q_s)+\frac{1}{4} M''(q_s)]
+(g_1+\frac{1}{2} g_2)V'''(q_s) + \frac{1}{8} V''''(q_s) < 0,
\label{cri}
\end{equation}
where $g_1$ and $g_2$ are evaluated at $\Omega=\Omega_O^{(1)}$.
 Since the dissipative coefficient is involved
at $\Omega^{(1)}$, $g_1$, and $g_2$, the criterion of first-order
phase transition is generally different from non-dissipative case.
Particularly, when particle mass is constant, the dissipation coefficient
 $\alpha$ appears only in $g_2$, and from the fact  $V''(q_s) < 0$
we can conclude that the Ohmic dissipation
enhances the possibility for the occurrence of the sharp first-order
transition.

\subsection{The super-Ohmic dissipation}
In this subsection we consider the criterion for the first-order phase
transition when the dissipation is super-Ohmic, i.e., 
$J(\omega)= \alpha \omega^3$, which appears in the analysis of macroscopic
magnetization tunneling\cite{ga89}.

 The dissipation term in Eq.(\ref{eqmotion}) with $n=3$ can be written
 again in terms of $\tilde{q}(\omega)$.
After integrating it  at complex $\tau'$-plane by using same prescription,
one can find the equation of motion (Eq.(\ref{eqmotion})) to be in the form
 \begin{equation}
M(q)\ddot{q} + \frac{1}{2}\frac{\partial M(q)}{\partial q} \dot{q}^2
+\alpha \int_{-\infty}^{\infty} d\omega
\tilde{q} (\omega) \mid \omega \mid^3  e^{i\omega \tau} = V'(q).
\label{eqmo2}
\end{equation}
In deriving Eq.(\ref{eqmo2}) we used 
 $ \delta V' = \frac{3\alpha}{2\epsilon} \ddot{q}$ to cancel a divergence
which appears in the course of integration.
The presence and absence of divergence in super-Ohmic and Ohmic cases, 
respectively, are because that we start with the effective action
(Eq.(\ref{effaction0})) as mentioned in the previous section.

The remaining calculation for super-Ohmic case is equivalent to that 
for the Ohmic case.
The differences are followings.
The $\hat{l}$ operator is changed in this case into
\begin{equation}
\hat{l} = M(q_s)\frac{d^2}{d\tau^2} + \frac{\alpha}{2\pi} \int d\omega
\mid \omega \mid^3  e^{i\omega \tau}  \int d\tau 
e^{-i\omega \tau},
\end{equation}
and the frequency near the sphaleron solution within the first order of
the amplitude $a$, $\Omega_S^{(1)}$, becomes the root of the equation
\begin{equation}
M(q)\Omega^2 - \alpha \mid \Omega \mid^3 + V'' =0.
\end{equation}
This equation tells that for the second-order phase transition the transition
temperature $T_c = \mid \Omega_S^{(1)} \mid/2\pi$ becomes higher than
that of the non-dissipative case, $\omega_s/2\pi$.
This means that contrary to the Ohmic dissipation case, the tunneling rate 
near $T_c$ is enhanced by the super-Ohmic dissipation. 

Next order calculation shows that $\eta_2$ has a same form with the previous
case(Eq.(\ref{etatwo})),
where $g_1$ is given at Eq.(\ref{g1g2}) and $g_2$ is changed in the form
\begin{equation}
g_2 = - \frac{3 \Omega^2 M'(q_s) + V'''(q_s)}
     {4[ 4 M(q_s)\Omega^2 - \alpha \mid 2 \Omega \mid^3 + V''(q_s)]}. 
\label{g2}
\end{equation}
Finally, the criterion of first-order phase transition is equivalent to
Eq.(\ref{cri}) except that $g_2$ is replaced by Eq.(\ref{g2}) and
 Eq.(\ref{cri}) should be evaluated at $\Omega=\Omega_S^{(1)}$
instead of $\Omega_O^{(1)}$.
It is, then, evident that when particle mass is constant the super-Ohmic
dissipation reduces the possibility for the occurrence of the sharp
first-order phase transition within exponential approximation.

\section{Application to quantum mechanical tunneling models}

\subsection{Asymmetric double well case}
Consider a usual double well potential with a symmetry-breaking term
proportional to $q^3$, i.e.,
\begin{equation}
V(q )= \frac{1}{2}(q^2 - 1)^2 -f q^3 .
\label{v3}
\end{equation}
This type of potential is frequently used in 
field theories for the application to cosmology\cite{ad93,wi98a}.
In this case the sphaleron solution is
\begin{equation}
q_s =0.
\end{equation}
We assume that particle mass $M$ is constant, so that the derivaltives of
mass are zero.
Substuting the derivatives of the potential at $q_s$
\begin{equation}
V''(q_s)=-2,\,\,\,\, V'''(q_s)=-6f,\,\,\,\, V''''(q_s)=12
\end{equation}
 and Eq.(\ref{g1g2}) into
Eq.(\ref{cri}), it is easy to obtain  following criterion
\begin{equation}
\frac{\alpha}{2}[\sqrt{(\frac{\alpha}{M})^2 + \frac{8}{M}}
-\frac{\alpha}{M}] > \frac{5}{2} + \frac{1}{2(3f^2+1)}
\end{equation}
for the first-order phase transition when the dissipation is Ohmic.
Unfortunately, for any values of positive $\alpha$ and $M$ the
left-hand side cannot exceed  2, i.e., always second order.
Hence, dissipation does not change the order of the phase transition.
This result is maintained even in the super-Ohmic case.

We can also obtain the criterion for a double well potential with
$q$ asymmetric term as
\begin{equation}
V(q )= \frac{1}{2}(q^2 - 1)^2 -F q .
\label{v1}
\end{equation}
In this case, the result is very similar to $q^3$ case.
There is no first-order phase transition.
This result can be expected since the both asymmetric potentials Eqs.(\ref{v3})
and (\ref{v1}),
when the dissipation term is ignored, give actions proportional to
each other, which can be easily shown by
a appropriate translation of $q$ and scalings of $q$ and $\tau$.

\subsection{ Case for spin tunneling-inspired model}

Consider an hamiltonian
\begin{equation}
H=\frac{p^2}{2M(\phi)} +V(\phi),
\label{modelh}
\end{equation}
where
\begin{equation}
M(\phi) = \frac{1}{2K_1 (1-\lambda \sin^2 \phi )},
\end{equation}
and
\begin{equation}
V(\phi) = K_2 S(S+1) \sin^2 \phi.
\end{equation}
Although this hamiltonian can be derived by the coherent representation
from the hamiltonian of spin tunneling model\cite{li98}
\begin{equation}
H=K_1 S_z^2+K_2 S_y^2
\label{spinh}
\end{equation}
where
$K_1$ and $K_2$ represent an anisotropic constants and $\lambda = K_2/K_1$,
 we will not go through the content of physics on spin tunneling in this
paper. Instead, we will restrict ourselves into the discussion on the
effect of dissipation in the hamiltonian, Eq.(\ref{modelh}).

Since the sphaleron solution is simply
\begin{equation}
\phi_s = \frac{\pi}{2},
\end{equation}
it is easy to obtain
\begin{eqnarray}
V''(\phi_s)&=&-2K_2 S(S+1),\, \,\,\,\, V'''(\phi_s)=0,
\,\,\,\,\, V''''(\phi_s)= 8 K_2S(S+1), \nonumber \\
M(\phi_s) &=& \frac{1}{2K_1 (1-\lambda )}, \,\,\,\,\, M'(\phi_s) =0,
\,\,\,\,\, M''(\phi_s) = - \frac{\lambda}{K_1}\frac{1}{(1-\lambda)^2}.
\end{eqnarray}
Hence the use of Eq.(\ref{cri}) for the Ohmic dissipation
 yields the condition for first-order phase
transition as
\begin{equation}
\frac{\alpha}{\sqrt{S(S+1)}} < \frac{2\lambda -1}{1-\lambda}.
\label{spincri}
\end{equation}
This result is shown in Fig.2.
It is shown that, unlike the case of constant mass, the effect of the Ohmic 
dissipation
reduces the range of parameters for the first-order phase transition.

Since it is impossible to get an analytic expression of criterion for the
super-Ohmic case, one has to resort to the numerical calculation.
The type of transition in the parameter space is given at Fig.3.
The effect of the super-Ohmic dissipation enlarges the parameter range for
the first-order phase transition, which is opposite behavior to the Ohmic case. 

It is worthwhile noting that the above criterion may not be realized in real spin
system. In order to study the effect of dissipation in the real spin system,
one has to derive the effective hamiltonian from the original
hamiltonian(Eq.(\ref{spinh})) with the dissipation term fully through
coherent representation or particle mapping\cite{za90}. Since this procedure
in general produces a complicated dissipation term, our cirterions
does not hold in this case. Therefore, what we have done in this subsection
is to obtain the general criterion for the sharp transition for the model hamiltonian
(Eq.(\ref{modelh})) with Ohmic and super-Ohmic dissipation and to show 
the possibility of the
change in type of phase tansition due to dissipation.

\section{conclusion}
Performing the nonlinear perturbation, the general condition for the sharp
first-order phase transition of decay rate between thermal regime and thermally
assisted quantum tunneling regime has been derived when position-dependent
 mass and dissipation
are involved. It has been shown that in the models with constant mass
 Ohmic dissipation enhances the possibility
for the occurrence of the sharp transition, while super-Ohmic dissipation 
reduces it.
Application to two simple quantum mechanical models is given at the previous
section.
In the asymmetric double well case the phase transition is always second order
regardless with or without dissipation. In this case, by comparing $\Omega_O^{(1)}$
and $\Omega_S^{(1)}$ with $\omega_s$ one can perceive that the Ohmic and super-Ohmic 
dissipation suppresses and enhances the decay rate, respectively,
 near the transition temperature $T_c$ within exponential approximation.
 Similar feature at zero temperature has been reported in Ref.\cite{fu92}.

For the case of $J(\omega)=\alpha \omega^2$ it is not clear how to give a proper
prescription which does not break the time-reversal symmetry. This difficulty
does not seem to originate from the property of Euclidean space.
In real time space,  a retarded
Green function is generally choosed in order to break the time-reversal symmetry.  
However, when $J(\omega)=\alpha \omega^2$, in spite of using the retarded Green function,
the time-reversal symmetry is not broken.
 
The generalization of our result, Eq.(\ref{cri}), to nonlinear coupling
between an environment and system and general type of dissipation might be
highly non-trivial. In this case the dissipation term in the equation of motion,
Eq.(\ref{eqmotion}), is proportional to
$$ \frac{\alpha}{\pi}\int_{-\infty}^{\infty} d\tau'
\frac{[q(\tau) - q(\tau')]^{z_2}}{\mid\tau-\tau'\mid^{z_1}} ,$$
where $z_1$ and $z_2$ are some constants.
Hence, the integration with respect to $\tau'$ through Fourier transform
as was done in Eq.(\ref{eqmo}) is impossible.
In our opinion the only way to break through this difficulty is to rely on
the numerical analysis. This will be discussed elsewhere.
It is also interesting to apply the present method to the spin tunneling
problems.
Since dissipation arises from the coupling with an environment which contains
quasi-particles like photons or phonons, it is possible to construct
a hamiltonian which contains the spin-environment coupling. Then, by
integrating over the environmental degrees of freedom and with the help of coherent
representation or particle mapping, one can derive an effective hamiltonian in which
the effect of dissipation is fully involved as a non-local term.
 From this procedure, one can investigate the
effect of the dissipation on the phase transition in spin tunneling problem.
This will also be discussed elsewhere.

\begin{center}
{\bf ACKNOWLEDGMENT}
\end{center}

S.Y. Lee acknowledges the financial support of KOSEF through a domestic
postdoctoral program. D.K. Park is grateful to D.A. Gorokhov for useful discussions.

\newpage
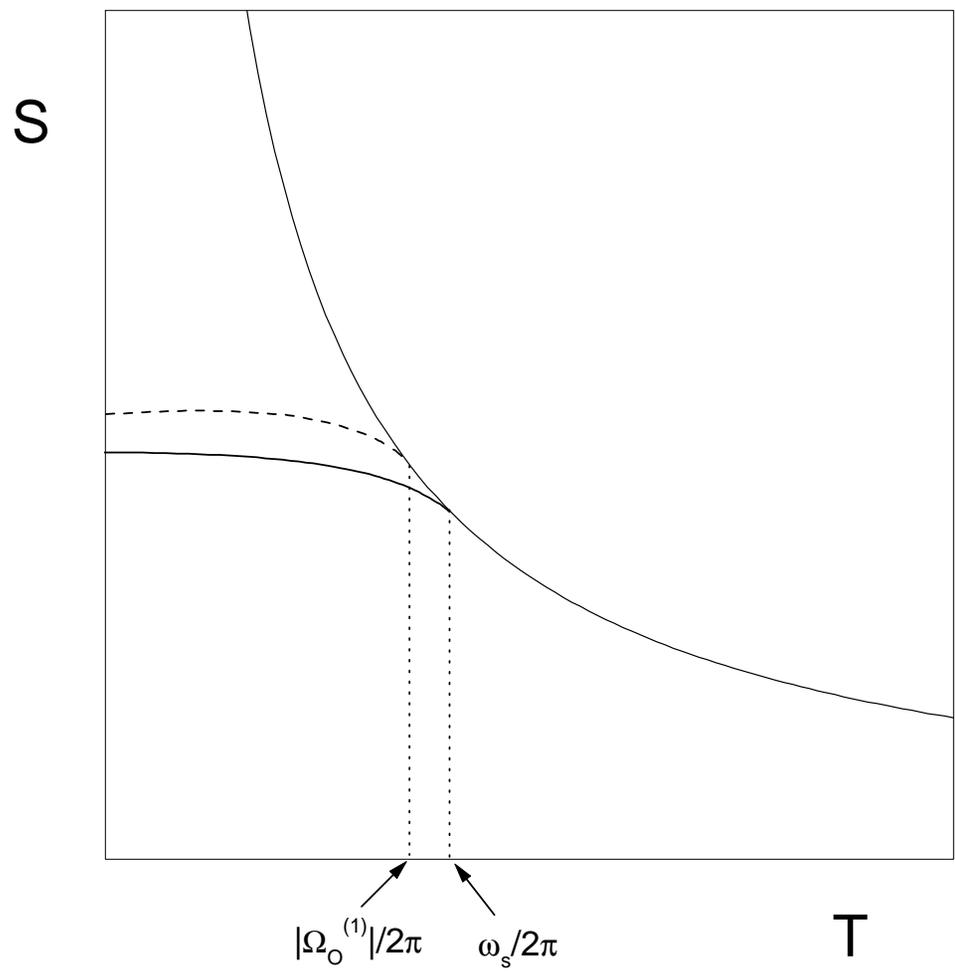
\begin{figure}
\caption{The seond-order phase transition. The decrease of transition 
temperature means the increase of action value near $T_c$, indicating suppression of
the decay rate. 
}
\end{figure}
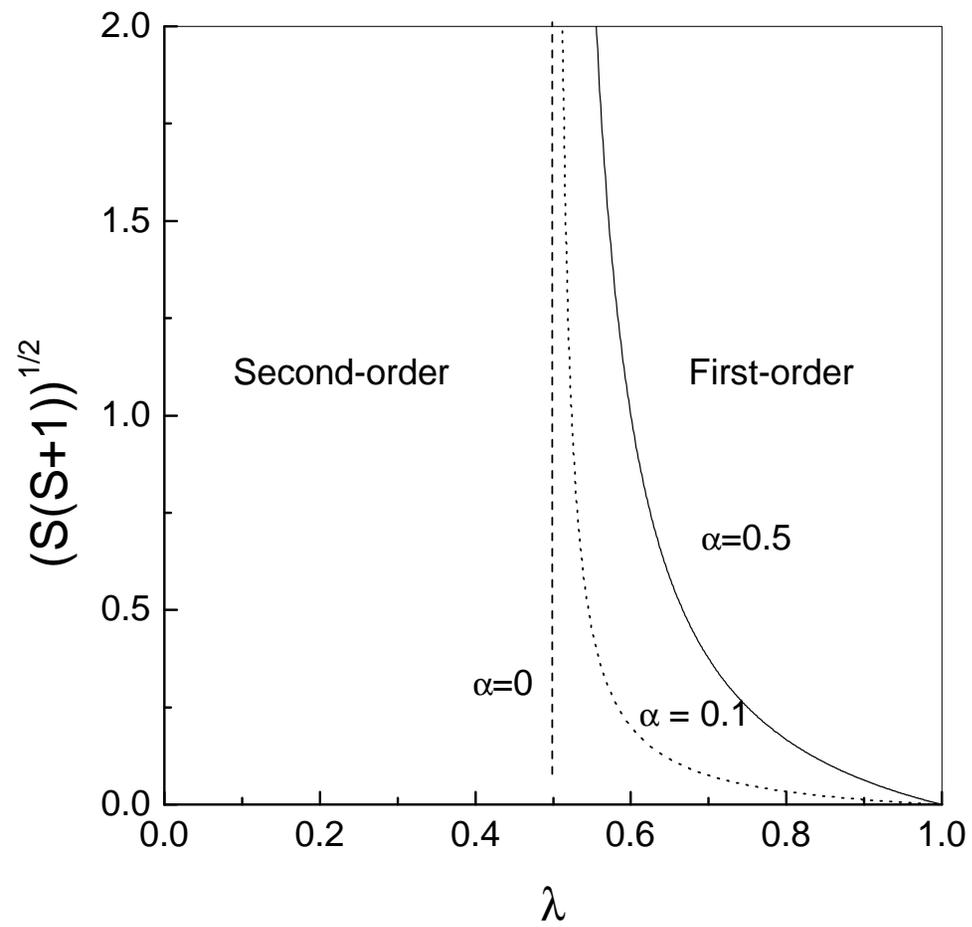
\begin{figure}
\caption{The parameter domains for first and second-order transitions
in the hamiltonian (Eq.(40)) with Ohmic dissipation.}
\end{figure}
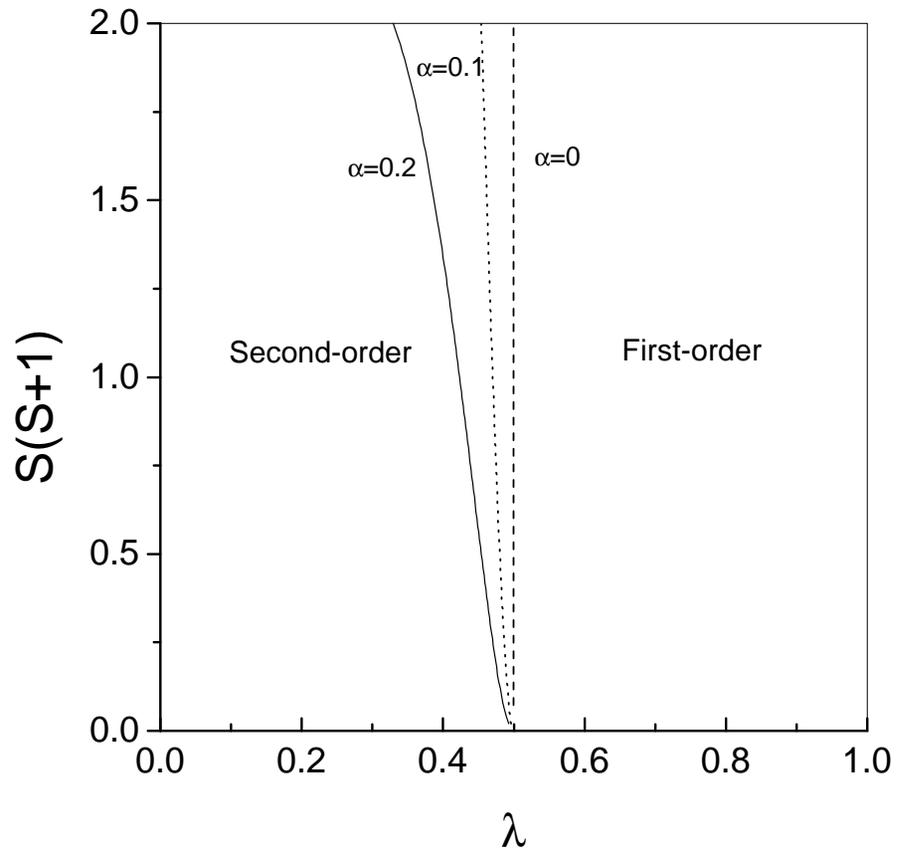
\begin{figure}
\caption{The parameter domains for first and second-order transitions
in the hamiltonian (Eq.(40)) with super-Ohmic dissipation.}
\end{figure}

\end{document}